\documentclass[12pt]{article}

\UseRawInputEncoding
\usepackage{amssymb,amsmath}
\usepackage{amsbsy}
\usepackage{amsthm}
\usepackage{mathrsfs}
\usepackage{verbatim}
\usepackage{mathdots}
\usepackage{graphicx,subfigure}
\usepackage[english]{babel}
\usepackage{enumerate}
\usepackage{caption}
\usepackage{pdfsync}
\usepackage{color,xcolor}
\usepackage{fullpage}
\usepackage{bbm}
\usepackage{comment}
\usepackage{authblk}

\usepackage{tikz}
\usetikzlibrary{backgrounds,fit, matrix}
\usetikzlibrary{positioning}
\usetikzlibrary{calc,through,chains}
\usetikzlibrary{arrows,shapes,snakes,automata, petri}

\newtheorem{Theorem}{Theorem}
\newtheorem{Corollary}{Corollary}
\newtheorem{Lemma}{Lemma}
\newtheorem{Proposition}{Proposition}

\theoremstyle{definition}

\newtheorem{Notation}[equation]{Notation}

\theoremstyle{remark}

\numberwithin{equation}{section}
\numberwithin{figure}{section}

\newcommand{\F}{{\mathbb F}}

\newcommand{\Z}{{\mathbb Z}}

\newcommand{\mbf}[1]{\mathbf{ #1}}

\begin{document}
\title{Error Correcting Codes From General Linear Groups}

\author[1]{Mahir Bilen Can}

\affil[1]{\small{
Tulane University, New Orleans, Louisiana\\
mahirbilencan@gmail.com}}

\maketitle

\begin{abstract}
The parameters of the AG codes on general linear groups are found.
The hyperplane sections having the minimum (or maximum) number of rational points are determined. 
\vspace{.5cm}

\noindent 
\textbf{Keywords: AG codes, good filtrations, hyperplane sections}

\noindent 
\textbf{MSC: 14G50, 94B27} 
\end{abstract}

\section{Introduction}

In this article, we analyze the hyperplane sections of general linear groups over finite fields.
The main goal of our paper is to investigate the simplest instance of a family of linear error correcting codes that we construct by using good filtrations. 
Let $GL_n(F)$ denote the general linear group of $n\times n$ matrices defined over a finite field $F$.
To motivate our discussion, we begin with the case of $2\times 2$ matrices over $F=\F_2$, the field with two elements. 
Let 
\begin{align}
GL_2(F):=
\left\{
\begin{bmatrix}
a & b \\
c & d
\end{bmatrix} :\
\{a,b,c,d\}\subseteq F\text{ and }\ ad-bc =1
\right\}.
\end{align}
The defining representation of $GL_2(F)$ is the two dimensional vector space $V:=F^2$. 
This is a simple $GL_2(F)$-module. 
Let $S(V^*)$ denote the symmetric algebra on the dual vector space $V^*$. 
By $S^r(V^*)$ we denote the homogeneous part of degree $r$ of $S(V^*)$. 
Every simple $GL_2(F)$-module is contained in one of the spaces $S^r(V^*)$ for some $r\in \{1,2,\dots\}$.
Let $M_2(F)$ denote the space of all $2\times 2$ matrices with entries from $F$. 
Then we can identify the tensor product $V_1:=S^1 (V^*)\otimes S^1(V)$ with the space of all homogeneous polynomial functions of degree 1 on $M_2(F)$.
The primary concern of our paper is the image, denoted by $C$, of the following ($F$-linear) evaluation map:
\begin{align}
ev_1 : V_1 & \longrightarrow F^{|GL_2(F)|} \notag \\
f &\longmapsto (f(A) | A\in GL_2(F)), \notag
\end{align}
where we use a fixed total order on the elements of the group $GL_2(F)$. 
After a straightforward computation, we see that 
\[
| V_1 | = 2^4 \ \text{ and }\ |GL_2(F)|=6. 
\]
Also, it is easy to check that 
\begin{align*}
C=
\left\{
\begin{array}{l}
(1,1,1,1,0,0), \  (1,0,0,0,1,0),\ (0,1,0,0,0,1),\ ( 0,0,1,0,1,0), \\
( 0,0,0,1,0,1),\ ( 0,0,1,1,1,1),\  (0,1,1,0,1,1),\ (1,0,0,1,1,1), \\
(1,1,0,0,1,1),\ (1,0,1,0,0,0),\ (0,1,0,1,0,0),\ (1,1,1,0,0,1),\\
(1,1,0,1,1,0),\ (1,0,1,1,0,1),\ (0,1,1,1,1,0),\ (0,0,0,0,0,0)
\end{array}
\right\}.
\end{align*}
By inspection we see that the minimum Hamming weight that is achieved by a nonzero vector in $C$ is 2.  
Therefore, the parameters of our code are given by $[6,4,2]_2$.
A code with these parameters is a subcode of the dual of the repetition code of length 6 over $F$.

Now, let $V$ denote the defining representation of $GL_n(F)$, where $n\geq 2$, and $F$ is a finite field with $q$ elements. 
Then $S^1 (V^*)\otimes S^1(V)$ can be identified with the space of all linear functionals on the space of all $n\times n$ matrices with 
entries from $F$. 
We denote by $C$ the algebraic geometry code that is obtained by evaluating the elements of $S^1 (V^*)\otimes S^1(V)$ on the points of $GL_n(F)$. 
Finally, we define, for a positive integer $r$, the $q$-analog of $r$ is defined by $[r]_q:=1+q+\cdots + q^{n-1}=\frac{q^n-1}{q-1}$.
As a convention, we set $[0]_q : = 1$. 
The factorial analog is now given by $[r]_q! := [r]_q [r-1]_q \cdots [2]_q [1]_q$.
In this notation, the first main theorem of our article is the following.

\begin{Theorem}\label{T:1}
Let $F=\F_q$. 
Let $C$ denote the evaluation code that is obtained by evaluating all homogeneous linear polynomials of the vector space $M_n(F)$ on the variety $GL_n(F)$.
Then the parameters of $C$ are given by 
\begin{enumerate}
\item $\mathtt{n}= q^{{n\choose 2}}(q-1)^n [n]_q!$,
\item $\mathtt{k}= {n^2}$,
\item $\mathtt{d}= q^{{n\choose 2}-1}(q-1)^{n-1} \left( (q-1)^2 [n]_q! - [n-2]_q!\right)$.
\end{enumerate}
\end{Theorem}

As a special case of this result we obtain the following corollary.

\begin{Corollary}\label{C:1}
Let $C$ be the code that is defined by evaluating linear forms on $GL_2(F)$, where $F= \F_q$.  
Then the length (denoted by $\mathtt{n}$), the dimension (denoted by $\mathtt{k}$), and the minimum distance (denoted by $\mathtt{d}$)
of $C$ are given by the following formulas  
\begin{enumerate}
\item $\mathtt{n}= q^4-q^3 - q^2 +q$,
\item $\mathtt{k}= {4}$,
\item $\mathtt{d}= q^4-2q^3 +q $.
\end{enumerate}
\end{Corollary}

\medskip

To prove Theorem~\ref{T:1}, hyperplane sections of $GL_n(F)$ with the greatest number of $F$-rational points are identified. 
Another goal of our paper is to show that the hyperplane sections of $GL_n(F)$ with the fewest $F$-rational points are derived from double cosets of Borel subgroups. 
\medskip

Let $B$ denote the Borel subgroup of all upper triangular matrices in $GL_n(F)$. 
The opposite Borel subgroup to $B$ is denoted by $B^-$. 
In other words, $B^-$ is the Borel subgroup of all lower triangular matrices in $GL_n(F)$. 
It is well-known that there are only finitely many $(B^-,B)$-double cosets in $GL_n(F)$.   
Chevalley's big cell theorem~\cite[Part II, \S1.9]{Jantzen} tells us that the geometric points of the double coset $B^-B$ is dense (in Zariski topology) in $GL_n(\bar{F})$, where $\bar{F}$ stands for an algebraic closure of $F$. 
The second main result of our paper identifies the complement of this set as a hyperplane section. 
\begin{Theorem}\label{T:2}
Let $Y$ denote the $(B^-,B)$-double coset $B^-B$ in $GL_n(F)$. 
Then $GL_n(F)\setminus Y$ is a hyperplane section with the fewest $F$-rational points in $GL_n(F)$.
\end{Theorem}

\medskip

The structure of our note is as follows. 
In the next section, we setup our notation. 
In Section~\ref{S:Lemmas} we observe the fact that the number of $F$-rational points of a general hyperplane section of $GL_n(F)$,
where $F$ is a finite field is equal to the number of $F$-rational points of a hyperplane section of $GL_n(F)$ that passes through the origin of $M_n(F)$.
In Section~\ref{S:Main}, we prove our first main result.
In Section~\ref{S:Main2}, we prove our second main result.
We conclude our paper by discussing the defects of our $GL_n(F)$-codes.

\section{Preliminaries and Notation}\label{S:Prelims}

The set of positive integers is denoted by $\Z_+$. 
For $n\in \Z_+$, the monoid of all $n\times n$ matrices with entries from a field $F$ is denoted by $M_n(F)$. 
\medskip

Let $F$ be a finite field with $q$ elements, where $q$ is a power of a prime number. 
If $X$ is an algebraic variety defined over $F$, then the notation $|X|$ will be used for denoting the number of $F$-rational points of $X$. 
When $q$ needs to be emphasized, we will use $\F_q$ instead of $F$.

The general linear group $G:=GL_n(\bar{F})$, where $\bar{F}$ is an algebraic closure of $F$, 
is an algebraic variety defined over $F$. 
The $F$-rational points of $G$ is given by the cardinality of $GL_n(F)$, which will be denoted by $\gamma(n,q)$. 
It is well-known that~\cite[Proposition 1.10.1]{EC1}, $\gamma(n,q) =  \prod_{i=0}^{n-1}  (q^n - q^i)$.
We write this product in the form $q^{{n\choose 2}}(q-1)^n [n]_q!$. 
The following recurrence follows at once: 
\[
\gamma(n,q) =  q^{n-1} (q^n-1) \gamma(n-1,q) \qquad \text{for $n\geq 2$}.
\]

\medskip

Since it is useful for our purposes, following~\cite{TVN}, we review a well-known reformulation of the $q$-ary linear codes.
An {\em $[\mathtt{n},\mathtt{k},\mathtt{d}]_q$ system} is a finite sequence $\mathcal{P}=(v_1,\dots, v_\mathtt{n})$, where 
$v_1,\dots, v_\mathtt{n}$ are not necessarily distinct vectors from an $F$-vector space $V$ such that 
\begin{itemize}
\item there is no hyperplane in $V$ that contains all of the vectors $v_1,\dots, v_\mathtt{n}$ (hence, $\mathtt{n}\geq \dim_{F}V$);
\item $\mathtt{k} =\dim_{F} V$;
\item $\mathtt{d} = \mathtt{n} - \max_H | \mathcal{P} \cap H|$,
where the maximum is taken over all hyperplanes $H\subseteq V$, and the points are counted with multiplicities. 
(It is assumed that $\mathtt{d}\geq 1$.) 
\end{itemize}
By~\cite[Proposition 1.1.4]{TVN}, there is a one-to-one correspondence $\varphi$ between the set of classes $[\mathtt{n},\mathtt{k},\mathtt{d}]_q$ systems and the set of linear $[\mathtt{n},\mathtt{k},\mathtt{d}]_q$ codes. 
Under this correspondence, $\varphi(\mathcal{P})$ is the code obtained by evaluating the linear functionals on $V$ at the entries of $\mathcal{P}$.

\section{Hyperplane Sections}\label{S:Lemmas}

Let $F$ be a field. 
Let $n\in \Z_+$. 
Hereafter, unless otherwise stated, we assume that $n\geq 2$. 
A {\em hyperplane} $H\subset M_n(F)$ is the shift by an element $C\in M_n(F)$ of the kernel of a linear map $B^\vee: M_n(F) \to F$.
This means that there exists a matrix $B\in M_n(F)$, and a scalar $c\in F$, such that 
\begin{align}\label{A:ourhyperplane}
H:=\{ A\in M_n(F) :\ \text{tr}(AB^\top)=c\}.
\end{align}
Note that, unless the scalar $c$ is zero, $H$ does not contain the zero of $M_n(F)$.
The purpose of this section is to analyze the cardinality of the intersection,  
\[
H \cap GL_n(F).
\]

\begin{Lemma}\label{L:intersectionwithGLn}
Let $B\in M_n(F)$ and $c\in F$. 
Let $H$ denote the hyperplane, $H:=\{ A\in M_n(F) :\ \text{tr}(AB^\top)=c\}$.
Then there exist $r\in [n]$, and two invertible matrices $E$ and $D$ in $GL_n(F)$ such that 
\begin{align*}
GL_n(F) \cap H &= \left\{ A\in GL_n(F) :\ \text{tr}\left( EAD^{-1}\begin{bmatrix}
\mbf{1}_r & 0 \\
0 & 0
\end{bmatrix} \right)=c\right\}.
\end{align*}
\end{Lemma}

\begin{proof}
We begin with noting that the trace form is invariant under conjugation by the elements of $GL_n(F)$:
\begin{align}\label{A:traceproperty}
\text{tr}( E C E^{-1} ) = \text{tr}(C)
\end{align}
for every $E\in GL_n(F)$ and $C\in M_n(F)$. 
We apply (\ref{A:traceproperty}) to the defining equation of $H$: 
\begin{align*}
H=\{ A\in M_n(F) :\ \text{tr}( EA B^\top E^{-1} )=c\}.
\end{align*}
Let $M_n^r(F)$ denote the set of all rank $r$ matrices from $M_n(F)$.
We proceed with the assumption that $B\in M_n^r(F)$. 
Then $B^\top$ is an element of $M_n^r(F)$ also.

The group $GL_n(F)\times GL_n(F)$ acts transitively on $M_n^r(F)$ by the left-right multiplication,
\begin{align*}
(D,E)\cdot A= DAE^{-1},\qquad\text{where}\qquad (D,E)\in GL_n(F)\times GL_n(F),\ A\in M_n^r(F).
\end{align*}
In other words, there exists $(D,E)\in GL_n(F)\times GL_n(F)$ such that
\begin{align*}
DB^\top E^{-1} = 
\begin{bmatrix}
\mbf{1}_r & 0 \\
0 & 0
\end{bmatrix},
\end{align*}
where $\mbf{1}_r$ is the $r\times r$ identity matrix.
Therefore, for $H$ as in (\ref{A:ourhyperplane}), there exist two invertible matrices $E$ and $D$ such that every element $A\in H$ satisfies 
\begin{align*}
\text{tr}\left( EAD^{-1}\begin{bmatrix}
\mbf{1}_r & 0 \\
0 & 0
\end{bmatrix} \right)=c.
\end{align*}
This finishes the proof of our assertion.
\end{proof}

\begin{Notation} 
Hereafter, we denote the rank $r$ idempotent matrix $\begin{bmatrix}
\mbf{1}_r & 0 \\
0 & 0
\end{bmatrix}$ by $e_r$.
\end{Notation}

We learn from Lemma~\ref{L:intersectionwithGLn} that not the matrix $B$ itself but its rank is more important.

\begin{Lemma}\label{L:2}
Let $H$ be a hyperplane of $M_n(F)$ as in the hypothesis of Lemma~\ref{L:intersectionwithGLn}.
Then there exists a hyperplane $H_0 \subset M_n(F)$ such that $\mbf{0} \in H_0$ and 
\[
|H\cap GL_n(F) | = | H_0 \cap GL_n(F)|. 
\]
\end{Lemma}

\begin{proof}
We follow the notation of the (proof of) Lemma~\ref{L:intersectionwithGLn}. 
Since the map $A\mapsto EAD^{-1}$, $A\in GL_n(F)$, is injective, 
we see that 
\begin{align*}
|GL_n(F)\cap H | = 
\left\vert \left\{ X\in GL_n(F) :\ \text{tr}\left( Xe_r \right)=c\right\} \right\vert.
\end{align*}
It remains to show that we can replace the scalar $c$ with 0. 
To this end, we will apply an affine transformation to $GL_n(F)$.

Let $C$ denote the matrix $-c e_r$. 
The map $\mbf{s}_C : M_n(F) \to M_n(F), \ Z\mapsto C+Z$ is an affine transformation of $M_n(F)$. 
It is easy to check that 
\begin{align}\label{A:aneasystep}
\left\vert \left\{ X\in GL_n(F) :\ \text{tr}\left( Xe_r \right)=c\right\} \right\vert 
&= 
\left\vert \left\{ X\in GL_n(F) :\ \text{tr}\left( \mbf{s}_C(X)e_r \right)=0\right\} \right\vert.
\end{align}
Notice that $\mbf{s}_C$ is a set-automorphism of $M_n(F)$. 
Let $\widetilde{GL_n}(F)$ denote the image of $GL_n(F)$ under $\mbf{s}_C$,
\[
\widetilde{GL_n}(F):= \{C+X :\ X\in GL_n(F)\}.
\]
It is easy to check that the product 
\begin{align*}
\widetilde{GL_n}(F) \times \widetilde{GL_n}(F) &\longrightarrow \widetilde{GL_n}(F) \\
(C+X, C+Y) &\longmapsto C+ XY
\end{align*}
makes $\widetilde{GL_n}(F)$ into a group that is isomorphic to $GL_n(F)$.

Let 
\begin{align*}
\varphi : \{ X\in GL_n(F) :\ \text{tr}\left( \mbf{s}_C(X)e_r \right)=0\} &\longrightarrow \{ \mbf{s}_C(X)\in \widetilde{GL_n}(F) :\ \text{tr}\left( \mbf{s}_C(X)e_r \right)=0 \} \\
X &\longmapsto \mbf{s}_C(X).
\end{align*}

Since $\mbf{s}_C(X)$ is a bijection, we see that $\varphi$ is a bijection. Hence, in light of (\ref{A:aneasystep}), we showed that 
\begin{align}\label{A:shifted}
\left\vert \left\{ X\in GL_n(F) :\ \text{tr}\left( Xe_r \right)=c\right\} \right\vert 
&= 
\left\vert \{ \mbf{s}_C(X)\in \widetilde{GL_n}(F) :\ \text{tr}\left( \mbf{s}_C(X)e_r \right)=0 \} \right\vert. 
\end{align}
Let $\widetilde{H}$ denote the hyperplane $\widetilde{H} = \{ \mbf{s}_C(Z) \in M_n(F) :\ \text{tr}( \mbf{s}_C(Z) e_r ) = 0 \}$. 
Then the right hand side of (\ref{A:shifted}) gives the cardinality of the intersection $\widetilde{GL_n}(F) \cap \widetilde{H}$.
In other words, we showed that $|\widetilde{GL_n}(F) \cap \widetilde{H}|= |GL_n(F) \cap H|$. 
Since $\mbf{0}\in \widetilde{H}$, and since $\widetilde{GL_n}(F) \cap \widetilde{H}$ is a hyperplane section of $\widetilde{GL}_n(F)$, the proof of our assertion is complete.
\end{proof}

\begin{Proposition}\label{P:intersectionwithGLn}
Let $H$ be a hyperplane section of $GL_n(F)$. 
Then, independent of the fact that $\mbf{0}\in H$ or not, there exists $r\in [n]$ such that 
\[
|H| =  \left\vert \left\{ \begin{bmatrix} A_{11} & A_{12} \\ A_{21} & A_{22} \end{bmatrix} \in GL_n(F) :\ 
\text{$A_{11}\in M_r(F)$ and $\text{tr}(A_{11}) = 0$}\right\} \right\vert.
\]
\end{Proposition}

\begin{proof}
By using Lemma~\ref{L:2}, we that there exists $r\in [n]$ such that 
\[
|GL_n(F) \cap H |= \vert \left\{ A\in GL_n(F) :\ \text{tr} (Ae_r)=0 \right\} \vert.
\]
After writing $A$ in the block form as in $A=\begin{bmatrix} A_{11} & A_{12} \\ A_{21} & A_{22} \end{bmatrix}$, where $A_{11}\in M_r(F)$
and $A_{22} \in M_{n-r}(F)$, we find that 
\[
|GL_n(F) \cap H| =  \left\vert \left\{ \begin{bmatrix} A_{11} & A_{12} \\ A_{21} & A_{22} \end{bmatrix} \in GL_n(F) :\ 
\text{$A_{11}\in M_r(F)$ and $\text{tr}(A_{11}) = 0$}\right\} \right\vert.
\]
This finishes the proof of our assertion.
\end{proof}

\section{Main Theorem}\label{S:Main}

We are now ready to prove our main theorem.
Let us recall its statement for convenience.
\medskip

Let $F=\F_q$. 
Let $C$ denote the evaluation code that is obtained by evaluating all homogeneous linear polynomials of the vector space $M_n(F)$ on the variety $GL_n(F)$.
Then the parameters of $C$ are given by 
\begin{enumerate}
\item $\mathtt{n}= q^{{n\choose 2}}(q-1)^n [n]_q!$,
\item $\mathtt{k}= {n^2}$,
\item $\mathtt{d}= q^{{n\choose 2}-1}(q-1)^{n-1} \left( (q-1)^2 [n]_q! -   [n-2]_q! \right)$.
\end{enumerate}

\begin{proof}

Recall that $\gamma(n,q)$ denotes the cardinality, $|GL_n(F)|= q^{{n\choose 2}}(q-1)^n [n]_q!$. 
This gives us the length $\mathtt{n}$. 
The dimension of our code is given by the dimension of the space of homogeneous linear forms on $M_n(F)$.
Clearly, this number is given by $n^2$. 
Finally, to find the minimum distance, we use Proposition~\ref{P:intersectionwithGLn} and a result of R. Stanley.
In his book~\cite[Chapter 1, Exercise 196 (a) and (b)]{EC1}, Stanley observed the following fact. 

For $k\in \{0,1,\dots, n\}$, let $f_k(n)$ denote the number of matrices $A= (a_{ij})\in GL_n(F)$ satisfying 
\[
a_{11}+a_{22}+\cdots + a_{kk} = 0.
\]
In other words, $f_k(n)$ is the number of $F$-rational points of the intersection of $GL_n(F)$ with the hyperplane $H\subset M_n(F)$ defined by 
the equation $a_{11}+a_{22}+\cdots + a_{kk} = 0$. 
Then, Stanley's formula is given by 
\begin{align}\label{A:Stanley's}
f_k(n) = \frac{1}{q} \left( \gamma(n,q) + (-1)^k (q-1) q^{\frac{1}{2} k(2n-k-1)} \gamma(n-k,q) \right).
\end{align}
Since $q$ and $n$ are fixed, we want to find $k_0$ such that 
\begin{align}\label{A:maxof}
f_{k_0}(n) = \max \{ f_k (n) : k\in [n] \}.
\end{align}
We claim that $k_0 = 2$. 

We split our proof into two parts. 
First we will show that $\frac{q}{q-1}(f_2(n)-f_1(n)) \geq 0$ for every $n$ and $q$ from $\{2,3,\dots\}$. 
Then we will show that $\frac{q}{q-1}(f_2(n) - f_j(n)) \geq 0$ for every $\{n,q\}\subset \{2,3,\dots \}$, and $j\in \{3,\dots, n\}$.

In the former case, we have 
\begin{align}\label{A:obvious1}
\frac{q}{q-1}(f_2(n)-f_1(n)) =  q^{2n-3} \gamma(n-2,q)  +  q^{n-1} \gamma(n-1,q). 
\end{align}
Evidently, the right hand side of (\ref{A:obvious1}) is nonnegative. 

In the latter case, we have 
\begin{align}\label{A:obvious2}
\frac{q}{q-1}(f_2(n)-f_j(n)) =   q^{2n-3} \gamma(n-2,q)  + (-1)^j   q^{\frac{1}{2} j(2n-j-1)} \gamma(n-j,q). 
\end{align}
In (\ref{A:obvious2}), similarly to the previous case, if $j$ is even, then there is nothing to do. 
We proceed with the assumption that $j$ is odd. 
Let $j= 2m+1$ for some $m\in \{1,\dots, \lfloor \frac{n}{2} \rfloor\}$.
Then we have 
\begin{align}\label{A:obvious22interim} 
\frac{q}{q-1}(f_2(n)-f_j(n)) &=   q^{2n-3} \gamma(n-2,q)  -  q^{ (2m+1)(n-m-1)} \gamma(n-2m-1,q).
\end{align}
Since $\gamma(n,q) = q^{{n\choose 2}}(q-1)^n [n]_q!$, we recognize the r.h.s. of (\ref{A:obvious22interim}) 
as follows:
\begin{align*} 
q^{2n-3+{n-2\choose 2}}(q-1)^{n-2} [n-2]_q!   -  q^{ (2m+1)(n-m-1)+{n-2m-1\choose 2}}(q-1)^{n-2m-1} [n-2m-1]_q!
\end{align*}
Since we have the inequalities, $(q-1)^{n-2m-1} \leq (q-1)^{n-2}$ and $[n-2m-1]_q! \leq [n-2]_q! $,
our task is reduced to a comparison of the terms  $q^{2n-3+{n-2\choose 2}}$ and $q^{ (2m+1)(n-m-1)+{n-2m-1\choose 2}}$. 
But it is a straightforward computation to show that 
\[
2n-3+{n-2\choose 2} = {n\choose 2}\ \text{ and }  (2m+1)(n-m-1)+{n-2m-1\choose 2} = {n\choose 2}.
\]
It follows that the inequality in (\ref{A:obvious22interim}), hence the inequality in (\ref{A:obvious2}) always hold true. 
This finishes the proof of our claim that $k_0 = 2$. 
Let $H_0\subset M_n(F)$ denote the hyperplane defined by $\{ (a_{ij})\in M_n(F):\ a_{11}+a_{22} = 0 \}$. 
Hence, the number of $F$-rational points of $H_0$ is given by $f_2(n)$. 
In light of this observation and Proposition~\ref{P:intersectionwithGLn}, we see that
\begin{align*}
\max_H | GL_n(F) \cap H|  = |H_0|,
\end{align*} 
where the maximum is taken over all hyperplanes of $M_n(F)$. 

We are now ready to prove our formula for $\mathtt{d}$:
\begin{align*}
\mathtt{d} &= \mathtt{n} - \max_H | GL_n(F) \cap H| \\
&= \mathtt{n} - f_2(n)\\
&= \left(1-\frac{1}{q} \right) \gamma(n,q) - \frac{1}{q} (q-1) q^{2n-3} \gamma(n-2,q) \\
&= q^{{n\choose 2}-1}(q-1)^{n+1} [n]_q! - q^{{n\choose 2}-1}(q-1)^{n-1} [n-2]_q!.
\end{align*}
By pulling $q^{{n\choose 2}-1}(q-1)^{n-1}$ out, we finish the proof of our theorem.
\end{proof}

\section{$B^-\times B$-stable Boundary}\label{S:Main2}

The goal of this section is to prove that the hyperplane section of $GL_n(F)$ with the minimal number of $F$-rational points is given by 
a $B^-\times B$ stable divisor of $GL_n(F)$, where $B$ stands for a Borel subgroup of $GL_n(F)$. 
To this end, we begin with setting up our algebraic group theoretic notation. 
The basic reference for this material is Jantzen's book, especially~\cite[Part II, \S1.9]{Jantzen}. 
\medskip

Let $F$ be a field.
Let $G$ be a (split) connected reductive algebraic group defined over $F$. 
We fix a (split) maximal torus $T$ in $G$, and we fix a Borel subgroup $B$ of $G$ such that $T\subseteq B$. 
The Weyl group of $(G,T)$ is denoted by $W$. 
It is given by $N_G(T)/T$, where $N_G(T)$ is the normalizer of $T$ in $G$. 
For $w\in W$, we denote by $\dot{w}$ a representative of $w$ in $N_G(T)$. 
The {\em Bruhat decomposition} of $G$ is the following partitioning of $G$ into double cosets of $B$:
\begin{align}\label{A:Bruhatdecomposition}
G = \bigsqcup_{w\in W} B \dot{w} B.
\end{align}
Then the Weyl group $W$ becomes a partially ordered set with respect to inclusion order on the Zariski closures of the double cosets: 
\begin{align*}
v \leq w \iff B\dot{v} B \subseteq \overline{B\dot{w}B}\qquad (w,v\in W).
\end{align*}
This order is called the {\em Bruhat-Chevalley order} on $W$. 
It is a graded poset with the rank function,
\begin{align*}
\ell : W & \longrightarrow \Z \\
w &\longmapsto \dim B\dot{w} B - \dim B.
\end{align*}
The value of an element $w\in W$ under $\ell$ is called the {\em length of $w$}. 
There is a unique maximal length element $w_0\in W$. 
It has the following pleasant properties:
\begin{itemize}
\item $\overline{B \dot{w_0} B} = G$, 
\item $w_0^2 = id$,  
\item $\dot{w_0} B \dot{w_0} \cap B = T$.
\end{itemize}
The Borel subgroup $\dot{w_0}B\dot{w_0}$, denoted by $B^-$, is called the {\em opposite} of $B$ (relative to $W$).
It is easy to see that Bruhat decomposition is equivalent to the following decomposition:
\begin{align}\label{A:Bruhat2}
G = \bigsqcup_{w\in W} B^- \dot{w} B.
\end{align}

We now focus on the most important linear algebraic group, namely, $G:=GL_n(F)$.
The group of all invertible upper triangular in $G$ is a Borel subgroup.
By abuse of notation we denote it by $B$. 
The group of invertible diagonal matrices in $G$ will be denoted by $T$.
Clearly, $T$ is a maximal torus contained in $B$. 
The Weyl group of $(G,T)$ is naturally isomorphic to the symmetric group $\mathbf{S}_n$. 
The opposite Borel subgroup of $B$, which is denoted by $B^-$, is the Borel subgroup of lower triangular matrices in $G$. 
Then the Bruhat decomposition $G = \bigsqcup_{w\in W} B^-wB$ is nothing but the 
LPU-decomposition of invertible matrices, where L stands for the lower triangular matrices, P stands for the permutations, and 
U stands for the upper triangular matrices. 
Note that we deliberately omitted the dots (on top of $w$) from our notation, since the symmetric group $\mathbf{S}_n$ is actually a subgroup of $G$.

Every Weyl group comes with a (not necessarily unique) system of Coxeter generators. 
In the case of the symmetric group $\mathbf{S}_n$ a Coxeter generating system is given by the set of simple transpositions
\begin{align}\label{A:Coxetergens}
S:= \{ s_i :\ i=1,\dots, n-1\},
\end{align}
where $s_i$ ($i\in \{1,\dots, n-1\}$) is the permutation of $\{1,\dots, n\}$ that interchanges $i$ and $i+1$ and leaves everything else fixed. 
Then the length $\ell(w)$ of an element $w\in \mathbf{S}_n$ is equal to the minimal number of simple transpositions that is need to 
express $w$ as their products. 
It follows from Matsumoto's exchange condition~\cite{Matsumoto} that, for every $s \in S$, we have 
\begin{align*}
\ell(w_0 s ) = \ell(w_0) -1\qquad\text{and}\qquad w_0 s \leq w_0.
\end{align*}
In other words, the set $\overline{B w_0 sB}$ ($s\in S$) is a one codimensional subvariety of $GL_n(F)$. 
We note in passing that 
\begin{align}\label{A:note}
\overline{B w_0 sB} = \overline{w_0 w_0 B w_0 sB} = w_0 \overline{B^- sB}.
\end{align}
Since $w_0$ acts transitively on $G$, we see that $\dim \overline{B w_0 sB} = \dim \overline{B^- sB}$ for every $s\in S$.
\medskip

\begin{Lemma}\label{L:H_0isSchubert}
Let $H_0$ denote the hyperplane $H_0 = \{ (a_{ij})_{i,j=1,\dots, n} \in M_n(F) :\ a_{11} = 0 \}$.
Then we have
\begin{align*}
H_0 \cap GL_n(F) = \bigsqcup_{s\in S} \overline{B^- s B},
\end{align*}
where $S$ denotes the set of all simple transpositions in $\mathbf{S}_n$. 
\end{Lemma}

\begin{proof}
By its definition, $H_0$ consists of all $n\times n$ matrices $x\in M_n(F)$ of the form 
\begin{align}\label{A:elementsofD}
x=
\begin{bmatrix}
0 & a_{12} & a_{13} & \cdots & a_{1n} \\
a_{21} & a_{22} & a_{23} & \cdots & a_{2n} \\
a_{31} & a_{32} & a_{33} & \cdots & a_{3n} \\
\vdots & \vdots & \vdots & \ddots & \vdots\\
a_{n1} & a_{n2} & a_{n3} & \cdots & a_{nn} \\
\end{bmatrix}.
\end{align}
By a straightforward matrix multiplication, we see that 
\begin{align*}
b^- x b \in H_0 \qquad\text{for every}\qquad (b^-,b)\in B^-\times B, \ x\in H_0. 
\end{align*}
In other words, $H_0$ is stable under the two-sided action of $B^-\times B$.
At the same time, by the Bruhat decomposition, $GL_n(F)$ is stable under the two sided action of $B^-\times B$ as well. 
Therefore, the intersection 
\[
\mathcal{D}:= H_0 \cap GL_n(F)
\] 
is a $B^-\times B$ stable subset of $GL_n(F)$. 
Since $H_0$ is hyperplane in $M_n(F)$, $\mathcal{D}$ is a hypersurface in $GL_n(F)$. 
We proceed to analyze its irreducible components. 
\medskip

Let $s\in S$. We claim that $\overline{B^- s B}$ is contained in $\mathcal{D}$. 
Clearly, to prove this claim, it suffices to show that $B^- s B\subseteq \mathcal{D}$. 
Let $b^-$ (resp. $b$) be an element from $B^-$ (resp. from $B$). 
It is easily checked that if $s= s_i$, where $i \in \{2,3,\dots, n-1\}$, then $b^- s b \in \mathcal{D}$.
We proceed to show that $b^- s_1 b \in \mathcal{D}$.
Here, it is harmless to assume that $n=2$. 
Then we have 
\begin{align*}
b^- s_1 b &=
\begin{bmatrix}
b_{11}' & 0 \\
b_{21}' & b_{22}'
\end{bmatrix}
\begin{bmatrix}
0 & 1 \\
1 & 0 
\end{bmatrix}
\begin{bmatrix}
b_{11} & b_{12} \\
0 & b_{22}
\end{bmatrix}\\
&=
\begin{bmatrix}
b_{11}' & 0 \\
b_{21}' & b_{22}'
\end{bmatrix}
\begin{bmatrix}
0 & b_{22}\\
b_{11} & b_{12}
\end{bmatrix}\\
&=
\begin{bmatrix}
0 & b_{11}'b_{22}\\
b_{11} b_{21}' & b_{21}'b_{22}+b_{22}'b_{12}
\end{bmatrix},
\end{align*}
which is an element of $\mathcal{D}$ also. 
In conclusion, we proved that 
\begin{align*}
\bigsqcup_{s\in S} \overline{B^- s B} \subseteq \mathcal{D}.
\end{align*}
It follows from (\ref{A:note}) that $\overline{B^-sB}$ is a $B^-\times B$ stable prime divisor in $GL_n(F)$. 
In fact, thanks to Bruhat decomposition, we know that every $B^-\times B$ stable prime divisor in $GL_n(F)$ is such an orbit closure for some $s\in S$. 
Since $\mathcal{D}$ is $B^-\times B$ stable, and since it contains every $\overline{B^-sB}$, where $s\in S$, 
we see that $\mathcal{D}$ is given by their union. This finishes the proof of our assertion. 
\end{proof}

We are now ready to prove the second main result of our article.

\begin{proof}[Proof of Theorem~\ref{T:2}]
Recall from the proof of Theorem~\ref{T:1} that  the number of $F$-rational points of the intersection of $GL_n(F)$ with the hyperplane 
$H = \{ A=(a_{ij}) \in M_n(F):\  a_{11} +\cdots a_{kk} = 0\}$ is given by 
\begin{align*}
f_k(n) = \frac{1}{q} \left( \gamma(n,q) + (-1)^k (q-1) q^{\frac{1}{2} k(2n-k-1)} \gamma(n-k,q) \right).
\end{align*}
We view $f_k(n)$ as a $\Z$-valued function. 
It is easy to check (along the lines of the proof of Theorem~\ref{T:1}) that the number $\min_{k\in [n]} f_k(n)$ is achieved by $f_1(n)$. 
Lemma~\ref{L:H_0isSchubert} shows that $f_1(n)$ is the number of $F$-rational points on the $B^-\times B$-stable divisor
$\bigsqcup_{s\in S} \overline{B^- s B}$.
But this divisor is precisely the complement of the ``Chevalley's big cell'' $B^-B$.
This finishes the proof of our theorem. 
\end{proof}

\section{Final Remarks}\label{S:Final}

Every linear $[\mathtt{n},\mathtt{k},\mathtt{d}]_q$ code satisfies the {\em Singleton bound}, that is the inequality $\mathtt{d} \leq \mathtt{n}-\mathtt{k} +1$. 
Likewise, every linear $[\mathtt{n},\mathtt{k},\mathtt{d}]_q$ code satisfies the {\em Griesmer bound}, that is the inequality $\sum_{i=0}^{\mathtt{k}}
\lceil \frac{\mathtt{d}}{q^i} \rceil \leq \mathtt{n}$. 
We call the differences $(\mathtt{n}-\mathtt{k} +1)- \mathtt{d}$ and $\mathtt{n} - \left(\sum_{i=0}^{\mathtt{k}}
\lceil \frac{\mathtt{d}}{q^i} \rceil\right)$ the  {\em Singleton defect} and the {\em Griesmer defect}, respectively. 
An {\em MDS code} is a linear code whose Singleton defect is 0. Such codes are among the most desirable codes. 
The $GL_n(F)$-codes are far from being an MDS code.
For $n=2$, we observe the following fact whose proof is straightforward. 
\begin{Proposition}
Let $F=\F_q$. 
Let $C$ denote the evaluation code that is obtained by evaluating all homogeneous linear polynomials of the vector space $M_2(F)$ on the variety $GL_2(F)$.
Then the following statements hold:
\begin{enumerate}
\item The Singleton defect of $C$ is given by $q^3-q^2 -3$. 
\item The Griesmer defect of $C$ is given by $q-1$. 
\end{enumerate}
\end{Proposition}

Over the binary field, we have the identification $GL_n(\F_2)=SL_n(\F_2)$.
Therefore, all of our results automatically hold true for $SL_n(\F_2)$.
For a non-binary field $F$, the parameters of the $SL_n(F)$-codes can be found in a similar way. 
In an upcoming manuscript we will present similar calculations for the other classical groups.

\section*{Acknowledgement}
This research was partially supported by a grant from the Louisiana Board of Regents (Contract no. LEQSF(2021-22)-ENH-DE-26).



\bibliography{references.bib}

\begin{thebibliography}{1}

\bibitem{Jantzen}
Jens~Carsten Jantzen.
\newblock {\em Representations of algebraic groups}, volume 107 of {\em
  Mathematical Surveys and Monographs}.
\newblock American Mathematical Society, Providence, RI, second edition, 2003.

\bibitem{Matsumoto}
Hideya Matsumoto.
\newblock G\'{e}n\'{e}rateurs et relations des groupes de {W}eyl
  g\'{e}n\'{e}ralis\'{e}s.
\newblock {\em C. R. Acad. Sci. Paris}, 258:3419--3422, 1964.

\bibitem{EC1}
Richard~P. Stanley.
\newblock {\em Enumerative combinatorics. {V}olume 1}, volume~49 of {\em
  Cambridge Studies in Advanced Mathematics}.
\newblock Cambridge University Press, Cambridge, second edition, 2012.

\bibitem{TVN}
Michael Tsfasman, Serge Vl\u{a}du\c{t}, and Dmitry Nogin.
\newblock {\em Algebraic geometric codes: basic notions}, volume 139 of {\em
  Mathematical Surveys and Monographs}.
\newblock American Mathematical Society, Providence, RI, 2007.

\end{thebibliography}
\bibliographystyle{plain}

\end{document}